\documentclass[sigconf]{acmart}

\usepackage{booktabs} 

\usepackage{booktabs}
\usepackage{amsmath}
\graphicspath{ {./images/} } 
\usepackage{subcaption}
\usepackage{multirow}
\usepackage{bm}

\definecolor{rewrite}{rgb}{1,0,0}

\usepackage{todonotes}

\setcopyright{rightsretained}

\copyrightyear{2018} 
\acmYear{2018} 
\setcopyright{acmcopyright}
\acmConference[KDD '18]{The 24th ACM SIGKDD International Conference on Knowledge Discovery & Data Mining}{August 19--23, 2018}{London, United Kingdom}
\acmBooktitle{KDD '18: The 24th ACM SIGKDD International Conference on Knowledge Discovery & Data Mining, August 19--23, 2018, London, United Kingdom}
\acmPrice{15.00}
\acmDOI{10.1145/3219819.3220065}
\acmISBN{978-1-4503-5552-0/18/08}

\begin{document}
\title[Identifying Sources and Sinks in the Presence of Multiple Agents with GP Vector Calculus]{Identifying Sources and Sinks in the Presence of Multiple Agents with Gaussian Process Vector Calculus}

\author{Adam D. Cobb}
\affiliation{
  \department{Department of Engineering Science}
  \institution{University of Oxford}  
  \city{Oxford}
  \state{United Kingdom}
}
\email{acobb@robots.ox.ac.uk}
\author{Richard Everett}
\affiliation{
  \department{Department of Engineering Science}
  \institution{University of Oxford}  
  \city{Oxford}
  \state{United Kingdom}
}
\email{richard@robots.ox.ac.uk}
\author{Andrew Markham}
\affiliation{
  \department{Department of Computer Science}
  \institution{University of Oxford}  
  \city{Oxford}
  \state{United Kingdom}
}
\email{andrew.markham@cs.ox.ac.uk}
\author{Stephen J. Roberts}
\affiliation{
  \department{Department of Engineering Science}
  \institution{University of Oxford}  
  \city{Oxford}
  \state{United Kingdom}
}
\email{sjrob@robots.ox.ac.uk}

\begin{abstract}
In systems of multiple agents, identifying the cause of observed agent dynamics is challenging. Often, these agents operate in diverse, non-stationary environments, where models rely on hand-crafted environment-specific features to infer influential regions in the system's surroundings. To overcome the limitations of these inflexible models, we present \textit{GP-LAPLACE}, a technique for locating sources and sinks from trajectories in time-varying fields. Using Gaussian processes, we jointly infer a spatio-temporal vector field, as well as canonical vector calculus operations on that field. Notably, we do this from only agent trajectories without requiring knowledge of the environment, and also obtain a metric for denoting the significance of inferred causal features in the environment by exploiting our probabilistic method. To evaluate our approach, we apply it to both synthetic and real-world GPS data, demonstrating the applicability of our technique in the presence of multiple agents, as well as its superiority over existing methods.
\end{abstract}

%
%



\maketitle

\section{Introduction}
Inferring the possible cause of an agent's behaviour from observing their dynamics is an important area of research across multiple domains \cite{albrecht2017autonomous}. For example, in ecology, the increasing availability of improved sensor technology and GPS data enables us to learn the motivations behind animal movement \cite{wilson2013locomotion,ellwood2017active}, assisting with animal conservation and environmental efforts. Other domains, such as robotics, use apprenticeship learning to construct reward functions to allow them to mimic their observations \cite{abbeel2004apprenticeship}.

Typical solutions for inferring the cause of an agent's behaviour tend to exploit Inverse Reinforcement Learning (IRL) \cite{russell1998learning} and include learning preference value functions \cite{chu2005preference} as well as utility, value, or policy surfaces from observed actions in a space. Unfortunately, these solutions often rely on the problem being easily framed as a Markov decision process \cite{puterman2014markov} which is not always appropriate. Instead, one may be interested in identifying an interpretable potential function, defined in continuous space, that can explain trajectories made by agents in a multi-agent system (MAS). This approach can be seen in previous work where agents are modelled as particles with their dynamics determined by a potential field \cite{brillinger2008three,brillinger2011modelling,preisler2013analyzing}.

In this paper, we consider observed trajectories influenced by a time- and space-varying potential function, and infer the spatio-temporal potential function from observed movement alone. To accomplish this, we present \textit{GP-LAPLACE}: Gaussian Processes for Locating Attractive PLACEs\footnote{Code and data: \url{https://github.com/AdamCobb/GP-LAPLACE}}. By constructing Gaussian processes (GPs) \cite{rasmussen2006gaussian}, our method jointly infers a spatio-temporal vector field as well as canonical vector calculus operations on that field, allowing us to estimate a time-varying map of sources and sinks of potential influence on agents. 

There are three notable advantages to our approach. First, it is able to reason about the interaction between agents and the environment from only agent trajectories, without requiring knowledge of the environment. Second, it allows the potential field to be non-stationary, which more accurately reflects the real-world. Finally, by exploiting a probabilistic method, we obtain a metric for denoting the significance of inferred causal features in the environment. 

To demonstrate the generality of our method as a tool for explaining agent and animal behaviour, we apply it to two distinct data sets. As an illustrative example, we evaluate our approach on a synthetic data set which we can compare to the true potential function (which isn't accessible in the real-world). Next, motivated by the existence of long-term GPS data, we apply our method to a real-world GPS data set of pelagic seabirds \cite{pollonara2015olfaction}, discovering a number of attractors which influence their behaviour. 

The rest of our paper is organised as follows: Section~\ref{sec:prelim} introduces the GP and its derivative manipulations for vector calculus, followed by Section~\ref{sec:model} where we present our model. In Section~\ref{sec:exp_syn} we evaluate our method on a synthetic data set where we know the true potential function. We then demonstrate how our approach can interpret real-world data by applying it to the Scolopi's shearwater GPS data set in Section~\ref{sec:exp_bird}. Finally, we highlight the novelty of our work compared to the existing literature in Section~\ref{sec:relatedwork} before concluding with future work in Section~\ref{sec:conclusion}.

\begin{figure}
  \centering
    \includegraphics[width=0.45\textwidth]{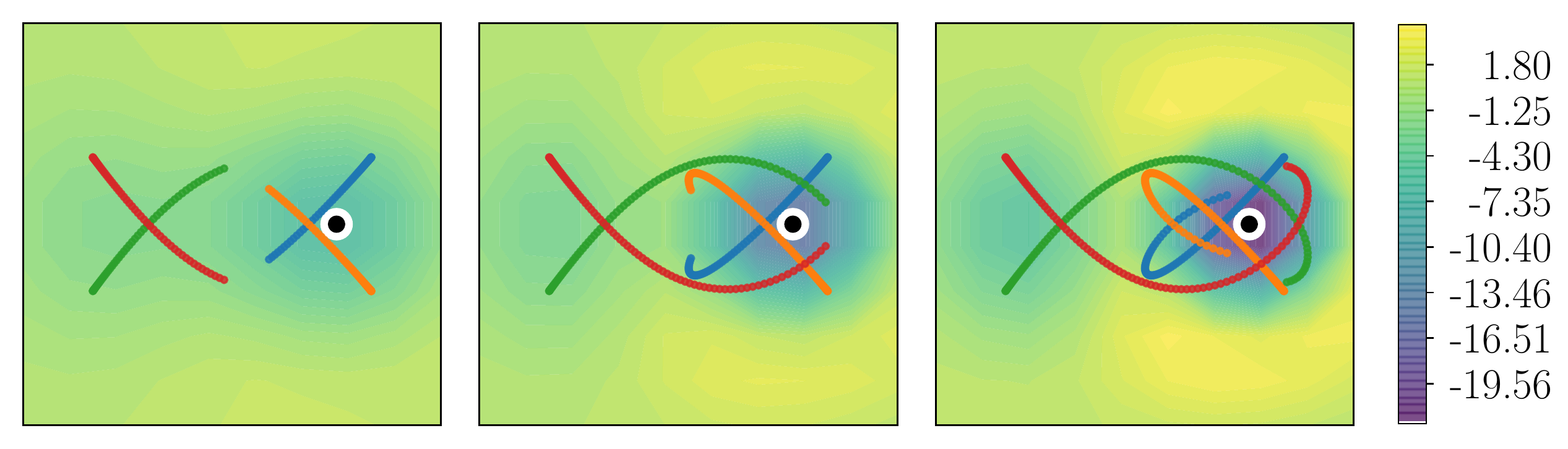}
    \caption{GP-LAPLACE tracking a time-varying attractor from observing four agent trajectories. The black and white marker indicates the true location of an attractor that is increasing in strength.}\label{fig:intro}
\end{figure}
     
\section{Preliminaries}
\label{sec:prelim}

As a requirement for our model, we introduce the Gaussian process, defined by its mean function $\boldsymbol{\mu}(\mathbf{x})$ and covariance function $\mathbf{K}(\mathbf{x,x'})$ \cite{rasmussen2006gaussian}.
The mean and covariance functions are parameterised by their hyperparameters and encode prior information into the model, such as smoothness, periodicity and any known underlying behaviour.
In our work, $\boldsymbol{\mu}(\mathbf{x})$ is set to zero as our data is preprocessed to have a zero mean. We define a function, distributed via a GP, as follows: 
\begin{equation}\label{eq:GP}
\mathbf{f(x)} \sim \mathcal{GP}\left(\boldsymbol{\mu}(\mathbf{x}),\mathbf{K}(\mathbf{x,x'})\right).
\end{equation}
Manipulating Gaussian identities using Bayes' rule gives formulae for the GP's posterior mean,
\begin{equation}\label{eq:mean_GP}
\mathbb{E}\left[f(\mathbf{x}_*)\right] = \mathbf{k}_{\mathbf{x}\mathbf{x}_*}^{\top}(\mathbf{K}_{\mathbf{x}\mathbf{x}}+\sigma^2\mathbf{I})^{-1} \mathbf{y},
\end{equation}
and posterior covariance,
\begin{equation}\label{eq:var_GP}
\mathbb{V}\left[f(\mathbf{x}_*)\right] = {k}_{\mathbf{x}_*\mathbf{x}_*} - \mathbf{k}_{\mathbf{x}\mathbf{x}_*}^{\top}(\mathbf{K}_{\mathbf{x}\mathbf{x}}+\sigma^2\mathbf{I})^{-1}\mathbf{k}_{\mathbf{x}\mathbf{x}_*},
\end{equation}
where $\mathbf{x_*}$ is a test point under question and $\sigma^2$ is the noise variance hyperparameter.

Any affine transformation of Gaussian distributed variables remain jointly Gaussian distributed.
As differentiation is an affine operation, applying this property to any collection of random variables distributed by a GP, gives jointly Gaussian distributed derivatives, $\mathbf{f'(x)}$. 
For a test point $\mathbf{x_*}$ and corresponding output $f(\mathbf{x}_*)$, the derivatives associated with the GP in Equation \eqref{eq:GP} are distributed with posterior mean,
\begin{equation}\label{eq:GP_der_mean}
\mathbb{E}\left[\frac{\partial^{n} f(\mathbf{x}_*)}{\partial\mathbf{x}_*^{n}} \right] = \frac{\partial^{n} \mathbf{k}_{\mathbf{x}\mathbf{x}_*}^{\top}}{\partial\mathbf{x}_*^{n}}(\mathbf{K}_{\mathbf{x}\mathbf{x}}+\sigma^2\mathbf{I})^{-1} \mathbf{y},
\end{equation}
and posterior covariance,
\begin{equation}\label{eq:GP_der_var}
\mathbb{V}\left[\frac{\partial^{n} f(\mathbf{x}_*)}{\partial\mathbf{x}_*^{n}} \right] = \frac{\partial^{2n} {k}_{\mathbf{x}_*\mathbf{x}_*}}{\partial\mathbf{x}_*^{n} \partial\mathbf{x}_*^{n}} - \frac{\partial^{n} \mathbf{k}_{\mathbf{x}\mathbf{x}_*}^{\top}}{\partial\mathbf{x}_*^{n}}(\mathbf{K}_{\mathbf{x}\mathbf{x}}+\sigma^2\mathbf{I})^{-1}\frac{\partial^{n} \mathbf{k}_{\mathbf{x}\mathbf{x}_*}}{\partial\mathbf{x}_*^{n}}.
\end{equation}
We define Equation \eqref{eq:GP_der_mean} as the predictive mean of the $n^{\text{th}}$ derivative with respect to any test points $\mathbf{x}_*$ and Equation \eqref{eq:GP_der_var} as its corresponding variance.

The choice of covariance selected throughout this paper is the squared exponential kernel,
\begin{equation}\label{eq:Kse}
k_{\text{SE}}(\mathbf{x}_* , \mathbf{x}_i) = l^2 \exp{\left(-\frac{1}{2}(\mathbf{x}_* - \mathbf{x}_i)^{\top}\boldsymbol{\Lambda}^{-1} (\mathbf{x}_* - \mathbf{x}_i)\right)},
\end{equation}
with $\mathbf{x}_*$ and $\mathbf{x}_i$ corresponding to a test point and training point respectively.
The hyperparameter  $\boldsymbol{\Lambda}$, is a diagonal matrix of input scale lengths, where each element determines the relevance of its corresponding dimension. The output scale, denoted by $l^2$, controls the magnitude of the output \cite{roberts2013gaussian}. The choice of kernel is motivated by the desire to obtain smooth measures over arbitrary derivative functions and the ease by which the kernel can be differentiated and used in combination with Equations \eqref{eq:GP_der_mean} and \eqref{eq:GP_der_var}.
The following formulae define the squared exponential kernel for the first and second order derivatives \cite{mchutchon2013differentiating}:
\begin{equation}\label{eq:Kse_1_der}
\frac{\partial k_{\text{SE}}(\mathbf{x}_*,\mathbf{x}_i)}{\partial \mathbf{x}_*} = - \boldsymbol{\Lambda}^{-1} (\mathbf{x}_* - \mathbf{x}_i) k_{\text{SE}}(\mathbf{x}_*,\mathbf{x}_i)
\end{equation}
\begin{equation}\label{eq:Kse_2_der}
\frac{\partial^2 k_{\text{SE}}(\mathbf{x}_*,\mathbf{x}_i)}{\partial \mathbf{x}_*^2} =  \boldsymbol{\Lambda}^{-1}\left( (\mathbf{x}_* - \mathbf{x}_i)(\mathbf{x}_* - \mathbf{x}_i)^\top\boldsymbol{\Lambda}^{-1} - \mathbf{I}\right) k_{\text{SE}}(\mathbf{x}_*,\mathbf{x}_i).
\end{equation}
We note at this point that estimating derivatives using a joint GP model over the function and derivatives \cite{brook2016emission,holsclaw2013gaussian} offers a benign noise escalation in comparison to numerical differentiation.

\subsection{Vector calculus with GPs}
\label{sec:vec_calc}
We define a scalar potential function $\phi(\mathbf{x},t)$ of space $\mathbf{x}$ and time $t$. Furthermore, we define the time-dependent gradient of $\phi(\mathbf{x},t)$ according to
\begin{equation}\label{eq:grad}
\boldsymbol{\nabla}_{\mathbf{x}}\ \phi(\mathbf{x},t) = \mathbf{V}_t
\end{equation}
with respect to $\mathbf{x}$, where $\mathbf{V}_t = \left[V_x\ V_y\right]^{\top}_t$ for $\mathbf{x} \in \mathbb{R}^2$.
We can model this time-varying vector value function by a multi-input, multi-output GP with a three-dimensional input tuple consisting of $\boldsymbol{\mathcal{X}} = (\mathbf{x},t)$.
This GP is constructed by introducing a separable kernel \cite{alvarez2012kernels}, such that
$$
\left[
\begin{array}{c}
V_x\\
V_y
\end{array}
\right]
\sim \mathcal{GP}\left(
\left[
\begin{array}{c}
\mu_x\\
\mu_y
\end{array}
\right],
\left[
\begin{array}{cc}
k_x(\boldsymbol{\mathcal{X}},\boldsymbol{\mathcal{X}}') & 0 \\
0 & k_y(\boldsymbol{\mathcal{X}},\boldsymbol{\mathcal{X}}')
\end{array}
\right]\right)
$$
contains an independently learned kernel for each output dimension.
Applying Equation \eqref{eq:GP_der_mean} to this GP model, by jointly inferring the derivatives in the $x$ and $y$ directions, gives the time dependent posterior for each random variable in the following tuple:
$$
\left( V_x,V_y,\frac{\partial V_x}{\partial x},\frac{\partial V_y}{\partial y}, \frac{\partial V_x}{\partial y},\frac{\partial V_y}{\partial x}\right)_t.
$$

We combine these predictive derivatives using Equations \eqref{eq:div} and \eqref{eq:curl} to infer probability distributions over the divergence and curl of the vector field\footnote{Vectors $\mathbf{\hat{i}},\mathbf{\hat{j}},\mathbf{\hat{k}}$ denote unit vectors of a 3-D Cartesian coordinate system.} $\mathbf{V}$:
\begin{equation}\label{eq:div}
\boldsymbol{\nabla}\cdot \mathbf{V} = \frac{\partial V_x}{\partial x} + \frac{\partial V_y}{\partial y},
\end{equation}
\begin{equation}\label{eq:curl}
\boldsymbol{\nabla}\times \mathbf{V} = \left| 
\begin{array}{ccc}
\mathbf{\hat{i}} & \mathbf{\hat{j}} & \mathbf{\hat{k}} \\
\frac{\partial}{\partial x} & \frac{\partial}{\partial y} & \frac{\partial}{\partial z} \\
V_x & V_y & 0 
\end{array} \right| =  \mathbf{\hat{k}}\left(\frac{\partial V_y}{\partial x} - \frac{\partial V_x}{\partial y}\right).
\end{equation}
Additionally, by simple application of the appropriate operators, we may readily define the time-varying spatial Laplacian:
\begin{equation}\label{eq:laplace}
\boldsymbol{\nabla} \cdot (\boldsymbol{\nabla} \phi) \equiv \boldsymbol{\nabla}^2 \phi.
\end{equation}
This (time-varying) Laplacian is of key importance as it defines \emph{sources} and \emph{sinks} in the spatial domain. It can be thought of as indicative of the flow in a vector field. Positive values denote attractive regions, where the vector field indicates flow towards these regions, whereas negative values denote repulsive regions due to the vector field pointing in the opposite direction.

\section{Model}\label{sec:model}

Our model builds upon the theory introduced in Section \ref{sec:prelim}.
The objective is to design a model that can indicate influential features in an agent's environment from observing their trajectories.

A trajectory $\boldsymbol{\zeta}_a$ for agent $a$ is defined as a collection of timestamped locations $\mathbf{x} \in \mathcal{R}^2$ or tuples $(\mathbf{x},t)$.
The elements of the vector $\mathbf{x}$ are referred to as $x$ and $y$ for the purposes of this model and we continue to use the tuple $\boldsymbol{\mathcal{X}} = (\mathbf{x},t)$ to refer to the domain of space and time.
We also make the assumption that each agent acts according to a utility or potential function $\phi(\mathbf{x},{t})$, which is dependent on space and time.
Whilst interacting with the environment, each agent aims to maximise this utility function at all times.

\subsection{Fitting to the agent trajectories}\label{sec:traj_fit}

The first component of GP-LAPLACE uses Equations \eqref{eq:mean_GP} and \eqref{eq:var_GP} to fit a GP to each trajectory $\boldsymbol{\zeta}_a$.
Using a single GP with a separable kernel, Equation \eqref{eq:joint_GP1} defines our GP prior over the $x$ and $y$ components of the path $\vec{\mathbf{f}}$:
\begin{equation}\label{eq:joint_GP1}
p(\vec{\mathbf{f}}) = \mathcal{GP}\left(\mathbf{0},\left[\begin{array}{cc}
k_x(t,t') & 0\\
0 & k_y(t,t')
\end{array}\right]\right).
\end{equation}
For each of the $x$ and $y$ components of $\vec{\mathbf{f}}$, a set of hyperparameters are learnt for the separable kernel.
The input space of this GP model is time $t$ and the output space consists of the $x$ and $y$ components of $\vec{\mathbf{f}}$. 
If an agent trajectory, $\boldsymbol{\zeta}_a$, consists of $N$ data points, we can then further apply Equation \eqref{eq:GP_der_mean} to infer higher order derivatives for each of the $n$ data points, where we denote $\{\dot{f}_x, \dot{f}_y\}$ and $\{\ddot{f}_x, \ddot{f}_y\}$ as the first and second-order time derivatives in the $x$ and $y$ directions.

Second-order derivatives are inferred at this stage, as we make the assumption that an agent acts in accordance with a second-order dynamical system, i.e. the agent obeys Newtonian dynamics.
This assumption means that an agent's underlying utility induces a ``force''  of influence on the agent, thus generating an acceleration (we here take the `influence-mass' of the agent as unity). More formally, this induces an acceleration equal to the derivative of the agent utility or potential:
\begin{equation}\label{eq:2nd_ass}
\ddot{f}_x = \frac{\partial \phi}{\partial {x}},\ \ \ddot{f}_y = \frac{\partial \phi}{\partial {y}}.
\end{equation}
Although we choose to infer second-order derivatives, the model is not limited to the assumption in Equation \eqref{eq:2nd_ass}. The flexibility of our model means that we can also infer first-order terms, $\dot{f}_x$ and $\dot{f}_y$, along with other higher-order terms. Therefore, throughout the rest of this section, references to $\ddot{f}_x$ and $\ddot{f}_y$ can be considered easily interchangeable with these other derivative terms.

When dealing with a multi-agent system of $M$ homogeneous agents, a trajectory model can be calculated for each agent to form the set of joint distributions,
$$\Big\{p(\vec{\mathbf{f}},\dot{f}_x,\dot{f}_y,\ddot{f}_x,\ddot{f}_y \mid \boldsymbol{\zeta}_a)\Big\}_{a=0}^M.$$
From the posterior GP model we are able to jointly predict the velocity and acceleration at any point on an agent's trajectory for $M$ agents. 
At this stage, we now have a collection of posterior derivatives and their corresponding location in $\boldsymbol{\mathcal{X}}$. If each of the agent trajectories has length $N$, then the size of   $\boldsymbol{\mathcal{X}}$ is $M\times N$.
The next layer of our model combines the outputs from the set of posterior distributions to construct a probability distribution over the extended agent environment.

\subsection{Inferring the vector field and Laplacian}\label{sec:inf_VF}
In order to infer the gradient of the potential function, $\boldsymbol{\nabla} \mathbf{\phi(\mathbf{x}},t)$, the set of inferred second derivatives for all $M$ agents is propagated through a second GP, which also has a separable kernel model, as below:

\begin{multline} \label{eq:joint_GP2}
p\Big(\vec{\mathbf{\mathbf{V}}}(\mathbf{x},t) \mid \{\ddot{f}_x,\ddot{f}_y,\boldsymbol{\zeta}_a\}_{a=0}^M\Big)\\ = \mathcal{GP}\left(\left[
\begin{array}{c}
\boldsymbol{\mu}_x^{\text{post}} \\
\boldsymbol{\mu}_y^{\text{post}}
\end{array}\right]
,\left[\begin{array}{cc}
k_x^{\text{post}}(\boldsymbol{\mathcal{X}},\boldsymbol{\mathcal{X}}') & 0\\
0 & k_y^{\text{post}}(\boldsymbol{\mathcal{X}},\boldsymbol{\mathcal{X}}')
\end{array}\right]\right).
\end{multline}
The vector $\vec{\mathbf{\mathbf{V}}}(\mathbf{x},t) = \left[{V}_x\ {V}_y\right]^{\top}$ consists of two random variables that model the acceleration in the two axes and the superscript label `$\text{post}$' refers to the calculated posterior mean and covariance. 
Equation \eqref{eq:joint_GP2} combines the $M$ multiple agent paths into one model and enables predictions to be made at different points in space that are not constrained to a single agent trajectory as in Equation \eqref{eq:joint_GP1}. The input-output pairs for this GP model are the $x,y$ and $t$ values in each $\boldsymbol{\zeta}_a$ that correspond to the $\ddot{f}_x$ and $\ddot{f}_y$ values.

The Newtonian assumption made in Section \ref{sec:traj_fit} is formally included as
$$
\boldsymbol{\nabla} \mathbf{\phi(\mathbf{x}},t) \propto \vec{\mathbf{\mathbf{V}}}(\mathbf{x},t).
$$
The distribution over the partial derivatives, $$\left[\frac{\partial V_x}{\partial x},\frac{\partial V_y}{\partial y}, \frac{\partial V_x}{\partial y},\frac{\partial V_y}{\partial x}\right],$$ can then be calculated from Equation \eqref{eq:joint_GP2} by applying Equations \eqref{eq:GP_der_mean} and \eqref{eq:GP_der_var}. We thus calculate a distribution over the divergence of $\vec{\mathbf{\mathbf{V}}}$. It follows that this divergence is proportional to the Laplacian under the same assumption,
\begin{equation}\label{eq:div_pot}
\boldsymbol{\nabla}^2 \mathbf{\phi(\mathbf{x}},t) \propto \boldsymbol{\nabla}\cdot \vec{\mathbf{\mathbf{V}}}(\mathbf{x},t).
\end{equation}

In particular, our interest lies in the estimation of the Laplacian of the utility function, as it indicates \emph{sources} and \emph{sinks} of the potential function in the environment. In this context, we regard \emph{sinks} as agent \emph{attractors} and \emph{sources} as agent \emph{repellers}. We have therefore introduced a novel framework, which enables us to infer sources and sinks, in an unsupervised manner, to offer an explanation behind multiple observed agent trajectories.

\subsection{Metric for locating significant attractors and repellers: Kullback--Leibler divergence} 
We now require a metric that is able to take advantage of having access to both the posterior mean and variance over sources and sinks in the environment.
Therefore, our metric of change from prior field to posterior field is measured via the Kullback--Leibler (KL) divergence \cite{kullback1951information}. The motivation for selecting the KL divergence comes from its ability measure a distance between two distributions. This provides a natural indication of the informativeness of spatial locations, at given times, and in the context of our application offers a measure of \emph{trajectory-influencing locations}.

Given the model at time $t$, each point in space has an associated potential field distribution, defined via the GP posterior as a univariate normal distribution.
The KL divergence can be readily calculated as the difference between two univariate normal distributions, namely the prior and posterior \cite{duchi2007derivations}, as below:
\begin{equation}\label{eq:KL_div}
D_{\text{KL}}(p_{\text{prior}} \mid \mid p_{\text{posterior}}) = \frac{1}{2}\left( \frac{\sigma_{\text{pr}}^2}{\sigma_{\text{po}}^2} + \frac{(\mu_{\text{po}}-\mu_{\text{pr}})^2}{\sigma_{\text{po}}^2} - 1 + \ln\left(\frac{\sigma_{\text{po}}}{\sigma_{\text{pr}}}\right)\right),
\end{equation}
where
$$
p_{\text{prior}} = \mathcal{N}(\mu_{\text{pr}},\sigma_{\text{pr}}^2), \ \ p_{\text{posterior}} = \mathcal{N}(\mu_{\text{po}},\sigma_{\text{po}}^2).
$$

We refer back to Equation \eqref{eq:div} and \eqref{eq:laplace} in order to calculate the following prior Laplacian at location $\boldsymbol{\mathcal{X}}_i$ in space-time:
$$
 \left(\frac{\partial V_x}{\partial x}+\frac{\partial V_y}{\partial y}\right) \sim \mathcal{N}\Big(0,\frac{\partial^2 k_x(\boldsymbol{\mathcal{X}}_i,\boldsymbol{\mathcal{X}}_i')}{\partial x^2}+\frac{\partial^2 k_y(\boldsymbol{\mathcal{X}}_i,\boldsymbol{\mathcal{X}}_i')}{\partial y^2}\Big),
$$
where, $
\frac{\partial^2 k_x(\boldsymbol{\mathcal{X}}_i,\boldsymbol{\mathcal{X}}_i')}{\partial x^2}+\frac{\partial^2 k_y(\boldsymbol{\mathcal{X}}_i,\boldsymbol{\mathcal{X}}_i')}{\partial y^2} =
\frac{h_x^2}{\lambda_x^2}+\frac{h_y^2}{\lambda_y^2}.
$ The hyperparameters $h_x$ and $\lambda_x$ are the output and input scale lengths of the $x$-part of the separable kernel in the GP model, with $h_y$ and $\lambda_y$ corresponding to the $y$-part.

As our interest lies in determining attractors and repellers in the field, a further addition to the KL divergence in Equation \eqref{eq:KL_div} is to multiply it by the sign of the posterior mean of the Laplacian. This multiplication carries over the prediction of negative sinks and positive sources, whilst measuring the divergence from the zero prior mean. We refer to this extension as the \emph{signed KL divergence}:
\begin{equation}
SD_{\text{KL}}(p_{\text{prior}} \mid \mid p_{\text{posterior}}) = \text{sign}(\mu_{\text{po}})\ D_{\text{KL}}(p_{\text{prior}} \mid \mid p_{\text{posterior}}).
\end{equation}
Large values in the $SD_{\text{KL}}$ indicate significant influential features in the agent environment, whereas small values around zero indicate that the model is likely to have reverted to its prior, the sensible prior being that there is no significant feature present.

\subsection{Computational complexity}
In order to use GP-LAPLACE on large data sets, we must overcome the computational complexity associated with inverting the $N\times N$ covariance matrix, which is $\mathcal{O}(N^3)$. 
There is a vast amount of literature that aims to overcome this issue, such as \cite{quinonero2005unifying}, and we use the sparse GP approximation built into the python package \texttt{gpflow} \cite{GPflow2017}. This approximation is based on work by \citet{titsias2009variational}, which we choose to implement when the total number of data points, $N$, exceeds $1000$. Using this sparse approximation allows our model to scale to large data sets, as will be shown in Section \ref{sec:exp_bird}.
\section{Application to synthetic data}\label{sec:exp_syn}

As an illustrative example, we apply GP-LAPLACE to synthetic data, where the true potential function and its derivatives are known. Importantly, this example allows us to evaluate the performance of our approach. We are then able to build on these results for the real-world data set in Section \ref{sec:exp_bird}, where we cannot possibly have access to the true potential function. 

This experiment consists of a multi-agent system of homogeneous agents, whose dynamics we observe. Our goal is to infer, from trajectories alone, the underlying potential value function.
We demonstrate that our approach is able to recover the potential field from a small number of agent trajectories by identifying the true sources and sinks. 
Therefore, we are able to recover trajectory-influencing locations in an unsupervised manner, with no prior knowledge of the environment.

\subsection{Agent model}\label{sec:ag_model}
Agents are modelled according to a second order dynamical system, whereby, at each time step $t$, the acceleration $\mathbf{\ddot{x}}$, velocity $\mathbf{\dot{x}}$ and position $\mathbf{x}$, with $\eta$ as the update increment, are given by:
$$\begin{array}{lcl}
\mathbf{\ddot{x}}_{t+1} & = & \boldsymbol{\nabla} \phi(\mathbf{x}_t,t),\\
\mathbf{\dot{x}}_{t+1} & = & \mathbf{\dot{x}}_t + \eta\ \mathbf{\ddot{x}}_{t+1},\\
\mathbf{x}_{t+1} & = & \mathbf{x}_t + \eta\ \mathbf{\dot{x}}_{t+1}.
\end{array}$$

\subsection{Agent potential function}
We define the causal potential function for our illustrative example as a Gaussian mixture model, where each Gaussian has both a time-varying mean $\boldsymbol{\mu}_i(t)$ and covariance $\boldsymbol{\Sigma}_i(t)$ that change harmonically according to:
\begin{equation}\label{eq:utility}
\phi(\mathbf{x},t) = \sum^K_{i=1}\alpha_i \mathcal{N}\big(\boldsymbol{\mu}_i(t), \boldsymbol{\Sigma}_i(t) \big),
\end{equation}
where $K$ defines the number of Gaussians and $\alpha$ is the weight of each corresponding Gaussian. For the purposes of our synthetic data, we can think of each Gaussian as a time-varying source or sink.

In order to validate our results, the first and second order derivatives of $\phi(\mathbf{x},t)$ are calculated using the known derivatives of a multivariate normal distribution $p(\mathbf{x})$ \cite{petersen2008matrix}:
\[
\frac{\partial p(\mathbf{x})}{\partial \mathbf{x}}=-p(\mathbf{x})\boldsymbol{\Sigma}^{-1}(\mathbf{x-m}),
\]
\[
\frac{\partial^2 p(\mathbf{x})}{\partial \mathbf{x} \partial \mathbf{x}^\top}=p(\mathbf{x})\Big(\boldsymbol{\Sigma}^{-1}(\mathbf{x-m})(\mathbf{x-m})^\top \boldsymbol{\Sigma}^{-1} - \boldsymbol{\Sigma}^{-1}\Big).
\]
These derivatives are then used in conjunction with the vector calculus operations in Equations \eqref{eq:div} and \eqref{eq:laplace} to calculate the true Laplacian of the utility function, which is used as a ground truth in our experiment.

\begin{figure}[h]
    \centering
    \begin{subfigure}[b]{.43\textwidth}
        \includegraphics[width=\textwidth]{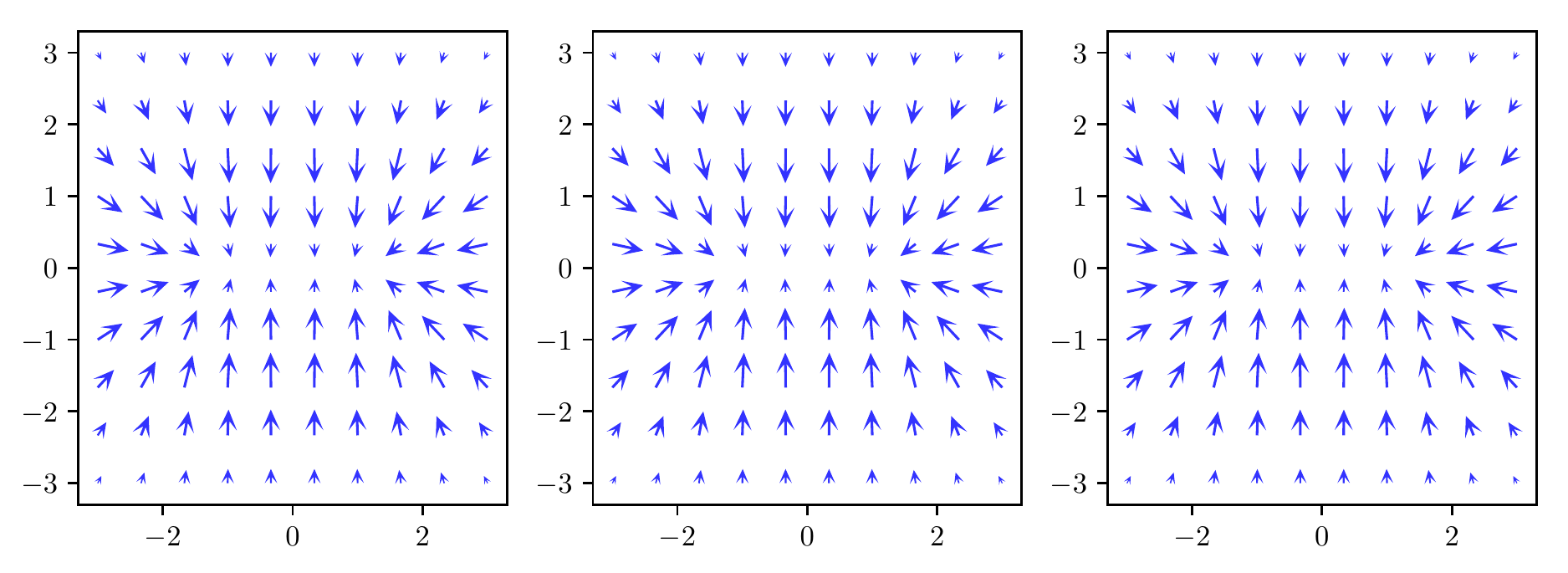}
     \caption{Stationary attractors}
     \label{fig:stat_field}
    \end{subfigure}
     \begin{subfigure}[b]{.43\textwidth}
        \includegraphics[width=\textwidth]{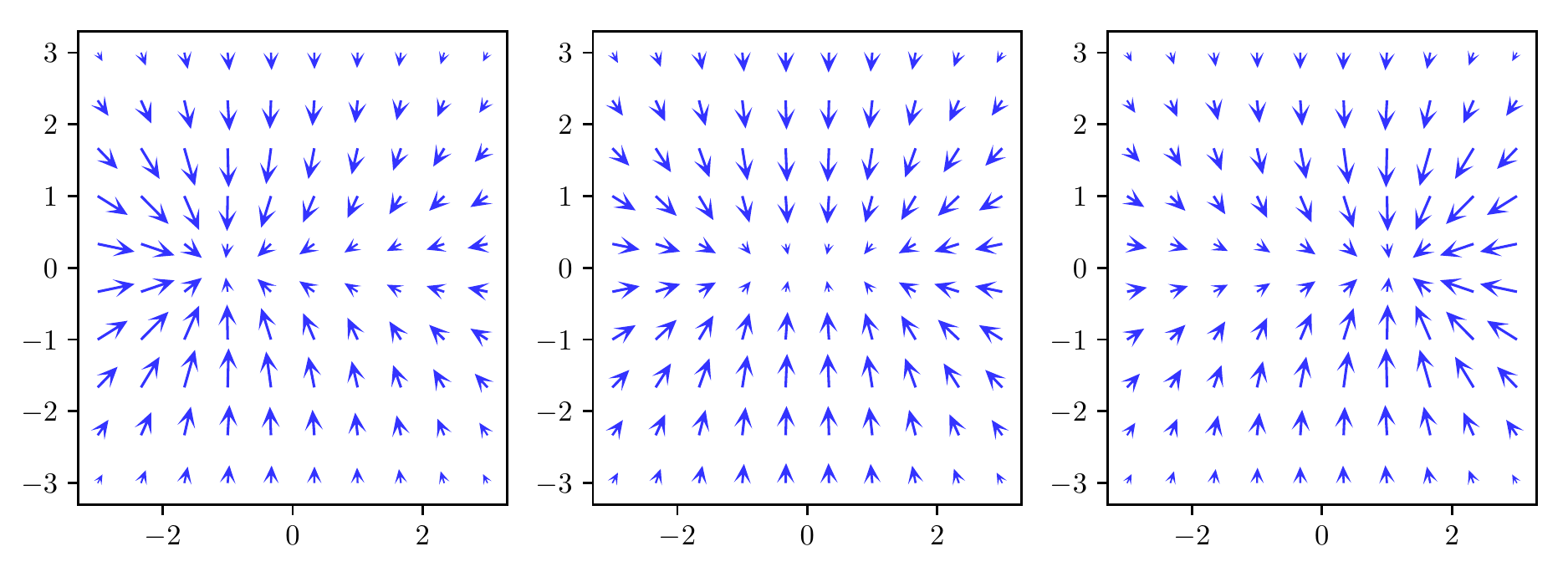}
       \caption{Varying-strength attractors} 
       \label{fig:var_field}
    \end{subfigure}
    \begin{subfigure}[b]{.43\textwidth}
        \includegraphics[width=\textwidth]{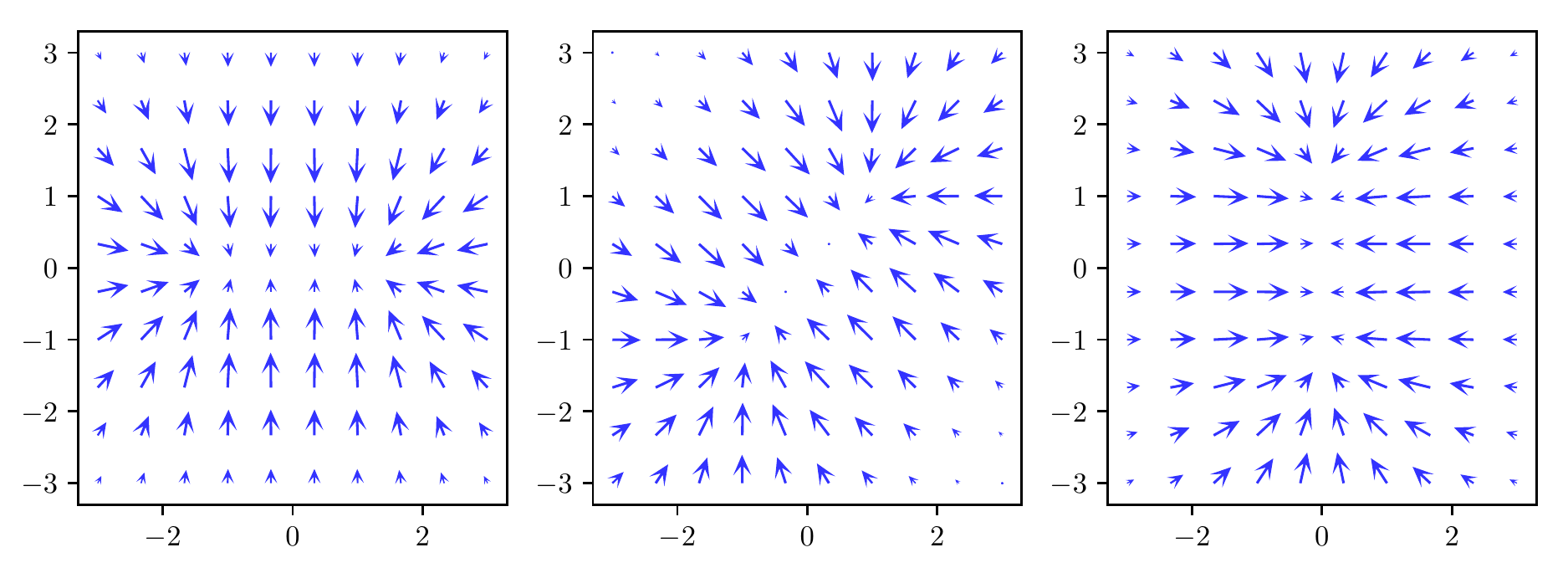}
        \caption{Rotating attractors}
        \label{fig:rot_field}
    \end{subfigure}
    \caption{We display the three example vector fields introduced in Section \ref{sec:ag_model}. Their time-varing properties are displayed by including three frames at different time-steps.}\label{fig:vec_fields}
\end{figure}

For the purposes of our experiment, we define three different agent potential functions:
\begin{enumerate}
\item \emph{stationary attractors} $$ \phi_1(\mathbf{x},t) = \sum^K_{i=1}\alpha_i \mathcal{N}\big(\mathbf{m}_i, c_i\mathbf{I} \big) $$
\item \emph{varying-strength attractors} $$ \phi_2(\mathbf{x},t) = \sum^K_{i=1}\alpha_i \mathcal{N}\big(\mathbf{m}_i, (\sin(t+\beta_i) + c_i)\mathbf{I} \big) $$
\item \emph{rotating attractors} $$ \phi_3(\mathbf{x},t) = \sum^K_{i=1}\alpha_i \mathcal{N}\big(\mathbf{m}_i \odot \left[ \begin{array}{c}
 \cos t\\
 \sin t
\end{array}\right], c_i\mathbf{I} \big) 
$$
\end{enumerate}
where $\alpha_i$, $\beta_i$ and $c_i$ are constants, $\mathbf{m}_i$ is a two-dimensional constant vector, and $\odot$ denotes an element-wise product. Here, we set $K=2$ for all three potential functions, along with enforcing $\alpha_i > 0$ to define attractors. Furthermore, we display the derivatives of these agent potential functions in Figure \ref{fig:vec_fields} to give an understanding of what they look like in practice.

Along with testing the performance of our model, we can also compare the vector fields in Figure \ref{fig:vec_fields} to those occurring in nature. As an example, a study by \citet{sommer2016information} tracked fruit bats as they moved between their camps and foraging sites. As nocturnal animals, their foraging sites are time-varying attractors, whereby these sites become the most attractive at night. Therefore, it is important to test our model on the time-varying attractors displayed in Figure \ref{fig:vec_fields}, where we have access to a ground truth, before applying our model to a real-world data set.

\subsection{Experimental results}\label{sec:syn_data}
For each of the three potential functions, we initialised $M$ agents with a velocity of zero at random starting locations. The experiment stepped through $200$ time-steps and our model is used to infer the vector field and Laplacian from the agent trajectories $\{\boldsymbol{\zeta}_a\}_{a=1}^M$ (see Equation \eqref{eq:div_pot}).

As a baseline, we took a simple parametric function in $x$- and $y$-space
\begin{equation} 
\begin{split}
\nabla U =&\   w_0 +w_1x +w_2y+w_3xy\ +w_4x^2+ \\
  &\ w_5y^2 +w_6x^2y+w_7xy^2+ w_8 x^3 +w_9 y^3
\end{split}
\end{equation}
of order three to model the gradient of the potential function, comparable to the approaches of \cite{brillinger2008three,brillinger2011modelling,preisler2013analyzing}.

Table \ref{tab:div_res} displays the results for the three experimental set-ups, where we have varied the number of $M$ observed agent trajectories from 4 to 16. We vary the number of agents to demonstrate how the models behave and scale as they observe more agents. Each value in the table is the mean squared error between the inferred Laplacian $\nabla^2 \tilde{\phi}$ and the true Laplacian $\nabla^2 \phi$, along with their standard deviation.

\begin{table*}
\caption{Results displaying the mean squared error between the true Laplacian and the inferred Laplacian. Each experiment consisted of 200 time-steps with the listed means and standard deviations calculated over 10 different random initialisations of the agents.}
  \begin{tabular}{l p{2cm} p{2cm} p{2cm} p{2cm}}
    \toprule
     &\multicolumn{4}{c}{Number of agents}\\  
    &4&8&12&16\\    
    \midrule
    \emph{Stationary attractors} - GP-LAPLACE & \bm{$6.38\pm2.84$} &  \bm{$2.20\pm2.81$} &  \bm{$0.55\pm0.27$} &  \bm{$0.35\pm0.39$}    \\
    \emph{Stationary attractors} - parametric &$8.27\pm0.84$ &  $8.22\pm0.48$ &  $8.02\pm0.13$ &  $7.99\pm0.49$\\
    \midrule
    \emph{Varying-strength attractors} - GP-LAPLACE & $62.37\pm99.9$ &  \bm{$0.88\pm0.37$} &  \bm{$0.64\pm0.11$} &  \bm{$0.58\pm0.14$}    \\
    \emph{Varying-strength attractors} - parametric & \bm{$23.59\pm7.23$} &  $19.15\pm3.37$ &  $17.67\pm1.95$ &  $19.19\pm2.59$    \\
    \midrule
    \emph{Rotating attractors} - GP-LAPLACE & $13.72\pm20.81$ &  \bm{$2.05\pm0.96$} &  \bm{$0.78\pm0.16$} &  \bm{$0.51\pm0.15$}    \\
    \emph{Rotating attractors} - parametric & \bm{$10.75\pm2.50$} &  $8.00\pm1.19$ &  $6.73\pm0.91$ &  $6.69\pm0.61$    \\
   
    \bottomrule
    \end{tabular}
    \label{tab:div_res}
\end{table*}

\begin{figure*}[h]
    \centering
    \begin{subfigure}[b]{.80\textwidth}
        \includegraphics[width=\textwidth]{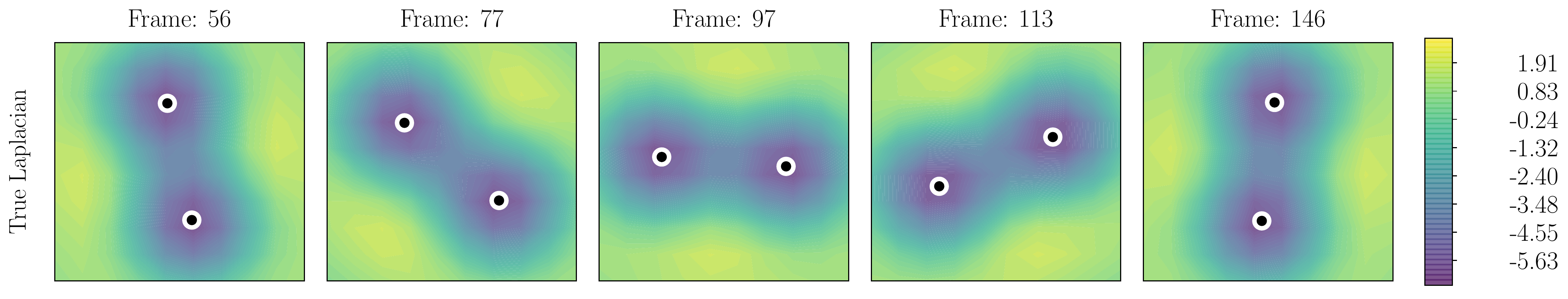}
        \label{skl_top}
    \end{subfigure}\\[-3ex]
     \begin{subfigure}[b]{.80\textwidth}
        \includegraphics[width=\textwidth]{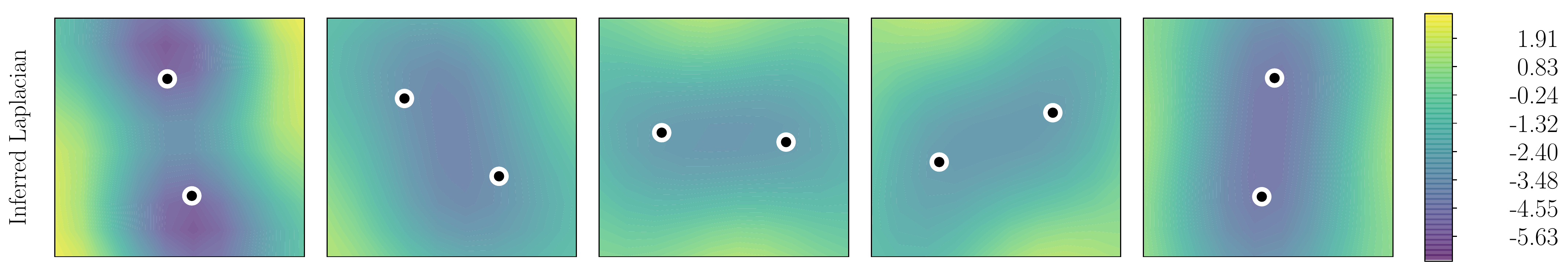}
        \label{skl_mid}
    \end{subfigure}\\[-3ex]
    \begin{subfigure}[b]{.80\textwidth}
        \includegraphics[width=\textwidth]{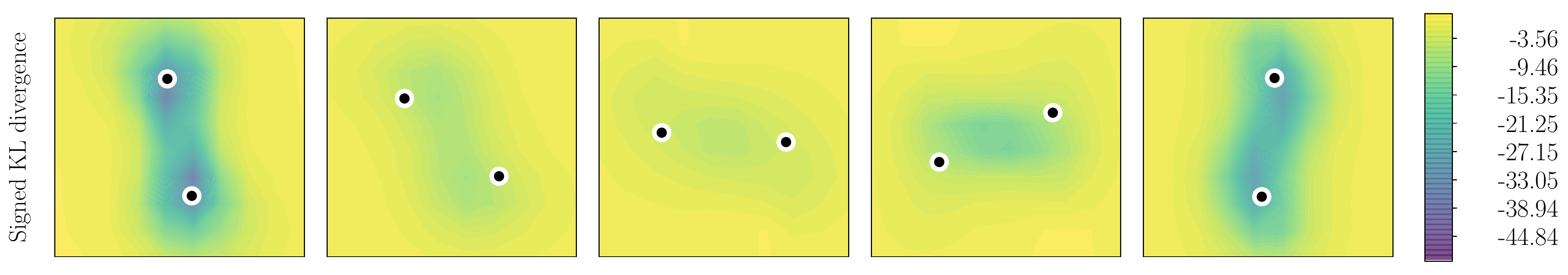}
        \label{skl_bot}
    \end{subfigure}
    \caption{Top row: Laplacian of true utility function. Middle row: inferred Laplacian of utility function. Bottom row: signed KL divergence. The location of the global minimum of the true Laplacian is indicated via the black and white markers. The predicted locations of the sinks, given by both the signed KL divergence and the posterior Laplacian, align well with their true locations, therefore demonstrating the success of our model on identifying non-stationary attractors.}\label{fig:frames_figures}
\end{figure*}

The results show that the parametric model was able to recover the general shape of $\nabla^2 \tilde{\phi}$ and often summarised the two attractors as a bowl shape in the three experiments. In comparison, GP-LAPLACE was able to recover the exact shape of the attractors for all three experiments and therefore demonstrated the ability to model the non-stationary characteristics of the potential functions.
A further result is that our model improved its accuracy as it observed more trajectories, which is a desirable property of the model, while the baseline did not consistently.

In order to interpret the behaviour of our model, the subset of frames displayed in Figure \ref{fig:frames_figures} show how the attractors of the true potential function are tracked across time by both the posterior inferred Laplacian and the signed KL divergence. This figure highlights how GP-LAPLACE is able to completely characterise \emph{rotating attractors} from only observing four trajectories over 200 time-steps. In more detail, the signed KL-divergence gives a measure of significance of the inferred sinks, through taking into account the uncertainty. The black and white markers indicate the location of the rotating attractors given by the ground truth in the first row.
For all three rows, regions of attraction are indicated by blue and it can be seen that the model is able to accurately track the true attractors across time.
\section{Application to Real-World Data}\label{sec:exp_bird}
In this experiment, we investigate how GP-LAPLACE can be used to determine driving forces in the environment that impact an animal's behaviour. We begin with an overview of the data used, present our experimental results, and then conclude with a discussion.

Unlike previous techniques for studying how animals interact with the environment, we don't rely on GPS data to build density maps to construct probability distributions of their likely locations \cite{horne2007analyzing,laver2008critical}. Furthermore, while more recent approaches have incorporated time into these models \cite{lyons2013home}, current methods do not focus on inferring the driving force behind animal actions and instead simply show where an animal is likely to be found. 

\subsection{Data}
We apply our model to a subset of Scolopi's shearwaters (Calonectris diomedea) \cite{sangster2012taxonomic} GPS data to infer the location of influential features in the environment of a MAS of pelagic seabirds. We use the same data set from Pollonara et al. (2015) \cite{pollonara2015olfaction}, made available in a \emph{Movebank data package} \cite{wikelski2014movebank}. 

In their experiment, shearwaters were released 400 km from the colony in the northern Mediterranean Sea. Using GPS trackers, the birds' trajectories were mapped and inferences were made about the way in which they navigated. Importantly, the birds were split into separate groups depending on which senses were inhibited. From these, we focus on two sets of birds consisting of the control set and the anosmic set, with the latter unable to use their sense of smell. 

Of note is that we do not constrain each trajectory to be equally spaced in time, nor do we require the GPS readings to be time aligned across the trajectories.

\subsection{Experimental Results}
\begin{figure*}[h]
    \centering
    \begin{subfigure}[b]{1.\textwidth}
        \includegraphics[width=\textwidth]{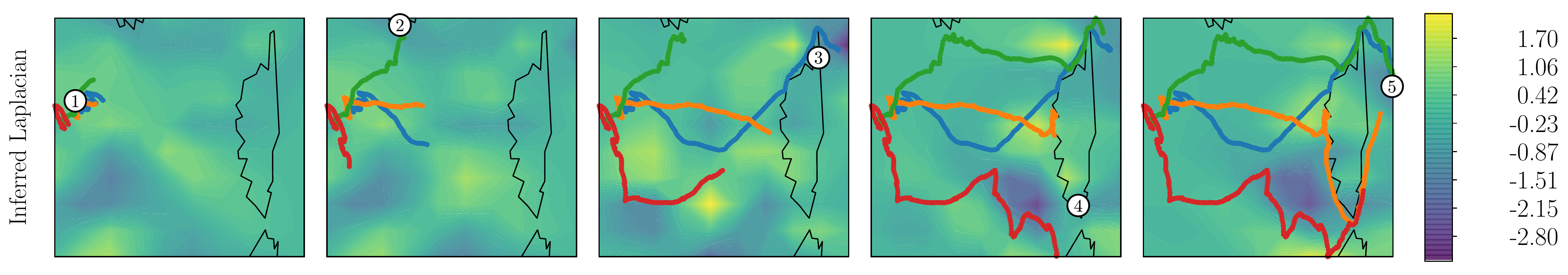}
        \label{bird_top}
    \end{subfigure}\\[-3ex]
     \begin{subfigure}[b]{1.\textwidth}
        \includegraphics[width=\textwidth]{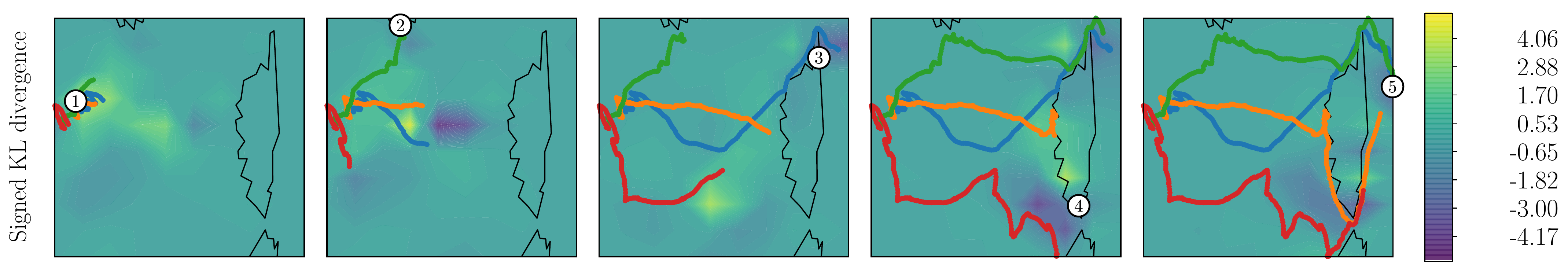}
        \label{bird_bot}
    \end{subfigure}\\[-3ex]
    \caption{Top row: Inferred Laplacian of utility function. Bottom row: Inferred signed KL divergence. Both rows are superimposed on a map of the northern Mediterranean Sea. Each frame represents a snapshot of the sources and sinks relating to the \emph{velocity flow} of the four shearwaters for 48 hours starting from 11:00 pm 21$^{\text{st}}$ June 2012. Points of note: (1) Release point, (2) South of France, (3) North of Corsica, (4) South of Corsica, (5) Nest. Best viewed in colour.}\label{fig:frames_figures_bird}
\end{figure*}

In Figure~\ref{fig:frames_figures_bird}, we present frames at different time-steps since the four control birds were released (in chronological order from left to right). Following previous work in this area \cite{brillinger2008three,brillinger2011modelling,preisler2013analyzing}, we focus on the divergence of the velocities of the birds. The top row is our inferred Laplacian, pointing out possible sources and sinks. The bottom row is the signed KL divergence, giving a measure of the significance of these features. 

Put into context, the birds' colony is slightly to the east of the northern part of Corsica, which is the island shown in the right side of each frame. Therefore, we expect to see our model placing an attractive region in the vicinity of their nest (marker (5) in figure), which is evident from the dark blue region appearing in the top-right of each of the last three frames, which is an attractor we expect to see as the birds start to approach their nest. Furthermore, the first frame shows that the region in which the birds are released is inferred to be a source as the birds fly outwards from this location (marker 1 in figure).

\begin{figure*}[h]
    \centering
        \includegraphics[width=1.0\textwidth]{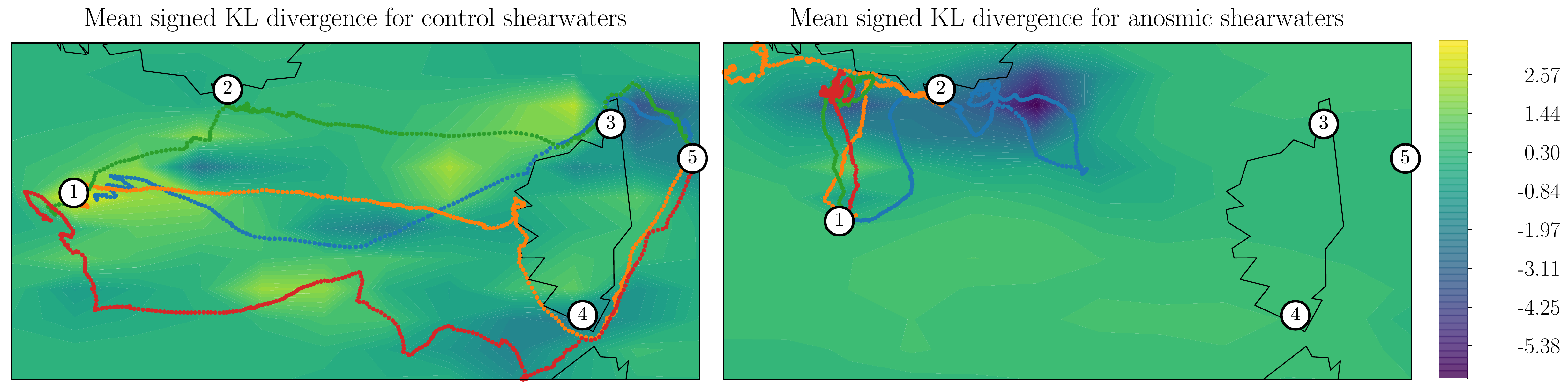}
        \caption{Left: Mean signed KL divergence for control shearwaters. Right: Mean signed KL divergence for anosmic shearwaters. Both plots contain trajectories of four separate shearwaters released at times similar times. Blue areas show attractive regions and the yellow denotes sources. The black outline corresponds to landmass, where the top is part of the southern French coastline and the island on the right is Corsica. Over a similar time period, the difference in behaviour can be seen from both the routes taken by the birds and the average placement of the sources and sinks. Points of note: (1) Release point, (2) South of France, (3) North of Corsica, (4) South of Corsica, (5) Nest. Best viewed in colour.}
        \label{fig:bird_comp}
\end{figure*}

Whereas the control birds in Figure \ref{fig:frames_figures_bird} tended to fly directly back to their colony, the anosmic birds flew north until they reached the southern coastline of France (marker (2) in figure). As pointed out in Pollonara et al. (2015) \cite{pollonara2015olfaction}, these birds are thought to have relied more on visual cues associated with the coastline, rather than flying straight back to their nest. Figure \ref{fig:bird_comp} shows the mean of the signed KL divergence across time for both the anosmic and control set of shearwaters, normalised to be on the same scale. 

In Figure \ref{fig:bird_comp}, a direct comparison can be seen between the two sets of birds. The left-hand side plot displays the mean across time for the control set of shearwaters, confirming that the bird colony is an attractor along with parts of the coastline of Corsica (marker (3) and (4) in figure). In contrast, the right-hand side plot for the anosmic shearwaters displays a different behaviour in both the distribution of the sources and sinks and in the trajectories themselves. As previously mentioned, the anosmic birds head North immediately after being released and use the coastline to recover their bearings (marker (1) and (2) in figure). Therefore our GP model, when taking the mean across time, clearly assigns the region along the coastline of southern France as an attractive region. Attributing the coastline as an attractive region to the anosmic birds agrees with the original suggestion by \citet{pollonara2015olfaction} that they navigate via `following coastlines as a form of search strategy, or by recognition of land features previously encountered'.

\subsection{Discussion}\label{sec:discussion}
A strength of our model is that there is no requirement to incorporate prior information on the location of environmental features. In our application to the shearwater GPS data set, the model had no prior knowledge of the location of the birds' colony or the coastline of Corsica, yet is still able to identify them as significant features. Furthermore, a strength of our technique is the way it is able to display how attractors and repellers vary with time. It would not be desirable to build a model that labels the birds' release point as a source for the entire duration of the birds' flight, although a simple density map would clearly make this error. As shown in the frames of Figure \ref{fig:frames_figures_bird}, GP-LAPLACE only describes their release point as a source for the first few hours of the birds' flights and the colony only becomes an attractor towards the end of the time period.

The results of this experiment mean that we are now able to take animal GPS data and infer which environment features are influencing their movements. We are not constrained by a requirement of modelling any details of the environment and therefore point to the generality of this technique as a flexible method that can be applied to many other data sets. The ability to model time-varying drivers in MASs points to further studies such as in \citet{ellwood2017active}, where we are interested in setting sampling rates for sensors and also making predictions as to which area of the environment will be attractive at different times of the day.

\section{Related work}\label{sec:relatedwork}
In this section, we refer to relevant work from both the GP literature and also from work relating to modelling GPS data.

As already introduced as a baseline in Section \ref{sec:syn_data}, there exists other works which model agents as acting according to a potential field \cite{brillinger2008three,brillinger2011modelling,preisler2013analyzing}. However, we show that GP-LAPLACE outperforms these methods due to the added flexibility of being able to model a time-varying potential function. There also exists alternative techniques for studying how animals interact with the environment, which often rely on using GPS data to build density maps to construct probability distributions of their likely locations \cite{horne2007analyzing,laver2008critical}. Although more recent approaches have incorporated time into these models \cite{lyons2013home}, they do not focus on inferring the dynamics and the driving forces behind animal actions, which is one of our motivations for introducing GP-LAPLACE.

When relating our use of GPs to previous work with vector fields, \citet{wahlstrom2013modeling} uses GPs to directly model magnetic fields. This work applies divergence-free and curl-free kernels to enforce linear constraints on their models, where the idea of constraining kernels to be divergence-free and curl-free is extended in \cite{jidling2017linearly}. Our work extends on previous work in this area with the introduction of a model that takes advantage of the derivative properties of GPs to give distributions over the operations of vector calculus from lower order observations. We have shown that inferring distributions over these operations results in an interpretable methodology that has not been previously employed in the literature. 

\section{Conclusion and Future Work}\label{sec:conclusion}
In this paper, we present GP-LAPLACE, a technique for locating attractors from trajectories in time-varying fields. Through applying Gaussian processes combined with vector calculus, we provide a model that is able to infer sources and sinks in the presence of multiple agents from their trajectories alone. Additionally, our probabilistic technique enables us to utilise the KL divergence to give a measure of significance of environmental features.

To demonstrate the generality of our method, we applied it to two data sets. First, on our synthetic data set, we showed that GP-LAPLACE is able to reconstruct non-stationary attractors effectively, with superior performance over the baseline parametric model. Next, in an unsupervised fashion, we applied our model to a real-world example where we were able to infer features of the environment, such as the release point and bird colony, without prior knowledge of the birds' surroundings.

In future work, we will extend GP-LAPLACE to incorporate explicit agent interactions. As an example, work by \citet{preisler2013analyzing} included an interaction term in their potential function to measure the strength of disturbances as having an effect on elk movement.
We will also look into applying our model to additional real-world data sets, in situations where discovering sources and sinks in the presence of multiple agents will offer further insight into the motivating factors behind agent behaviour.

\begin{acks}
Adam D. Cobb is sponsored by the AIMS CDT (\url{http://aims.robots.ox.ac.uk}) and the EPSRC (\url{https://www.epsrc.ac.uk}). Richard Everett is sponsored by the UK EPSRC CDT in Cyber Security. We also thank Ivan Kiskin for extensive comments and feedback. 
\end{acks}

\bibliographystyle{ACM-Reference-Format}
\bibliography{sample-bibliography}

\end{document}